%% file: wtup_mont10.tex
\documentclass[final,3p,times,twocolumn]{elsarticle}
 \biboptions{comma,sort&compress}
\usepackage{here}
 \usepackage{graphicx}
  \usepackage{epsfig}
\def\nin{\noindent}
\def\beq{\begin{equation}}
\def\eeq{\end{equation}}
\def\bea{\begin{eqnarray}}
\def\eea{\end{eqnarray}}

\def\pythia{{\sc Pythia}}
\def\herwig{{\sc Herwig}}
\def\CASCADE{{\sc Cascade}}
\def\PYTHIA{{\sc Pythia}}
\def\lesssim{\ \hbox{\raise 2pt \hbox{$<$} \kern -13pt
                     \lower 3pt \hbox{$\sim$}}\ }
\def\greatersim{\ \hbox{\raise 2pt \hbox{$>$} \kern -13pt
                     \lower 3pt \hbox{$\sim$}}\ }
\begin{document}

\begin{frontmatter}



\title{Forward jets in  hadron-hadron  collisions }

% \author[label1]{F. Hautmann\corref{cor1}}
% \address[label1]{Department of Theoretical Physics, University of Oxford, 
%Oxford OX1 3NP  }
%\cortext[cor1]{Speaker}
%\ead{snarison@yahoo.fr}

% \author[label1,label2]{Marina Nielsen,\corref{label3}}
%  \address[label2]{Instituto de F\'{\i}sica, Universidade de S\~{a}o Paulo, 
%C.P. 66318, 05389-970 S\~{a}o Paulo, SP, Brazil}
%\cortext[label3]{Supported by FAPESP within the France-Brazil program.}
%\ead{mnielsen@if.usp.br}
\author{F. Hautmann}

\address{Department of Theoretical Physics,  University   of Oxford,  
Oxford OX1 3NP }

\begin{abstract}
\noindent
This talk  discusses  basic  aspects of forward  production of jets   in 
pp collisions at high energy, including i)  issues on 
  QCD  factorization for hard processes at large rapidities, and ii)  the role of 
forward jet measurements at the LHC to investigate  contributions to parton showers 
from large-angle  gluon radiation  and   from multiple parton  interactions. 
\end{abstract}

%\begin{keyword}


%\end{keyword}

\end{frontmatter}

% \linenumbers

\section{Introduction}
%\label{}
\nin
%%%%%%%%%%%%
Physics in the forward region at hadron colliders 
is traditionally dominated by soft particle production. With the  start  of the LHC, 
forward physics  turns  into  a  largely new field~\cite{grothe,ajaltouni,denterria}        
because,  
due to the phase space opening up at large center-of-mass energies, 
both soft and hard production processes become relevant  and, 
 thanks to the unprecedented reach in rapidity 
 of  the experimental instrumentation,  
it becomes possible to carry out a program of  jet 
 physics in the forward region.  
Hard processes at forward rapidities    enter  
the LHC physics program   in  an  essential 
way    both for QCD studies and for new particle searches, e.g. 
in  vector boson fusion search channels for  the 
Higgs boson~\cite{vbf-cms,vbf-atlas}.

Forward production of  high  p$_{\rm{T}}$  brings jet physics into a 
  region  characterized  by  multiple energy scales and   
asymmetric parton kinematics.     
It   has long been recognized that 
 reliable theory predictions  in this region require  
  the  resummation of high-energy QCD corrections~\cite{muenav,webetal99}. 
 For the LHC forward jet kinematics, 
  QCD  logarithmic  corrections    in the large rapidity  interval   (of  high-energy type)  
and   in  
the hard transverse momentum (of collinear type)
may both be quantitatively  significant~\cite{ajaltouni}.     The theoretical framework 
to resum   consistently 
both  kinds of logarithmic corrections   to all  perturbative orders    
   is based on QCD  high-energy    factorization   
 at  fixed transverse momentum~\cite{hef}.  
This  factorization program is carried through  in~\cite{jhep09}  
for forward    jet hadroproduction. 
We discuss this briefly  in Sec.~2.

Besides these  different types of radiative corrections to 
single parton scattering,  the need for     
realistic  Monte Carlo simulations     of 
 forward particle production  also  raises the question of whether  
non-negligible effects may come from 
  multiple parton interactions~\cite{bartfano}.  Such multiple 
 interactions are modeled in  parton-shower   event generators 
used to simulate  final states  at the 
LHC~\cite{pz_perugia,rdf1,Sjostrand:2006za,giese}, and form the 
subject of a number of   current 
efforts~\cite{blok,strik-vogel,rog-strik,calu-trel,wiede_mpi,maina_mpi,berger_mpi,gaunt} 
to construct  approaches  that  incorporate 
  multiple parton scatterings.

\begin{figure}[htb]
\vspace{45mm}
\includegraphics{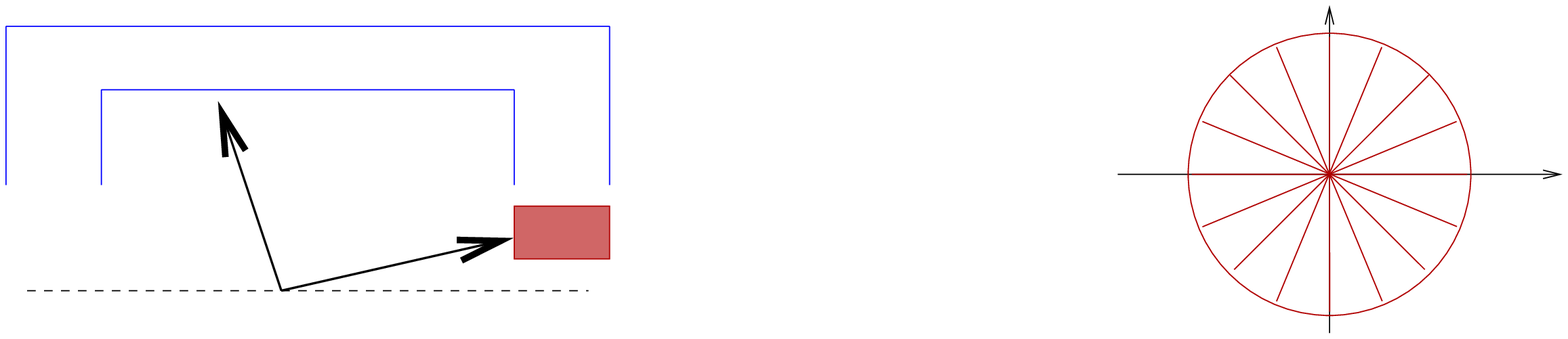}
\includegraphics{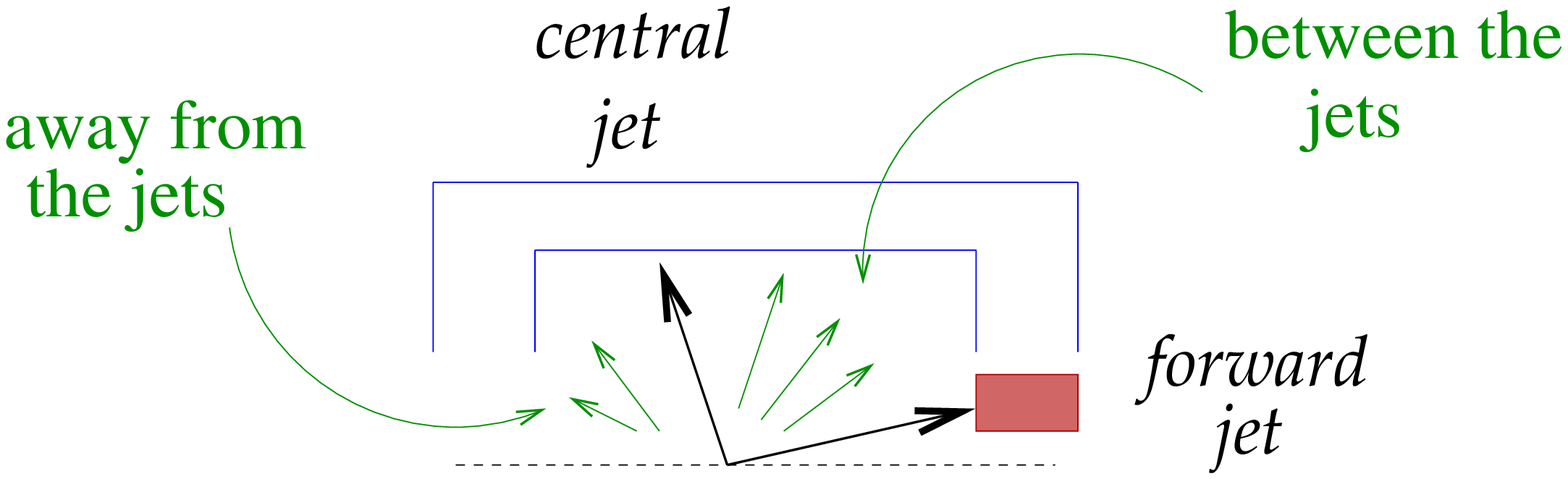}  
\caption{(top) Jets in the forward and  central detectors, and 
azimuthal plane segmentation;   
 (bottom) particle and energy flow in the inter-jet and outside regions.} 
\label{fig:jetcorr} 
\end{figure}

The capabilities  of forward + central detectors at the LHC 
suggest  the possibility to make a combined phenomenological study of 
multi-parton interactions versus 
 higher-order radiative contributions to single-parton interaction 
by examining correlations of one forward and one central jet~\cite{preprint} 
in  rapidity and azimuth (Fig.~\ref{fig:jetcorr}).   
We show   results   on  this  in Sec.~3. 

Because  forward jet  production probes  the gluon  density function   
for small x, it can   naturally be used 
 to investigate  possible   
 nonlinear  effects~\cite{ianmue} 
 at  high parton density.  The formulation~\cite{jhep09} at fixed transverse momentum   
 is   well-suited for  describing  the approach to the  high-density  region, as 
 it is designed to take
  into account  both the  effects from  BFKL  evolution associated with the 
  increase in rapidity  and  also  the 
 effects from   increasing    p$_{\rm{T}}$
 described by  renormalization group, which are  found to be also quantitatively 
 significant~\cite{kov10}  
  for studies of  parton saturation.  See 
  e.g.~\cite{kutak-absorpt} for  
 first    Monte Carlo  calculations along these lines, 
 and~\cite{gelis_etal_rvw} for extension 
 to nucleus-nucleus collisions.

Many of the 
  theoretical  issues  that underlie  forward jet  physics, from  
   perturbative QCD resummations to    approaching   the saturation region to 
   parton-showering methods beyond leading order,      
 depend on   the notion of transverse momentum dependent, or unintegrated, 
parton distribution functions (u-pdfs).   (See~\cite{hj_rec} for  recent reviews on this 
topic.)    
In the calculations presented below   we take the   high-energy 
 definition of u-pdfs~\cite{hef}, namely, 
we  rely on  the fact that  for small x 
  u-pdfs can be defined 
gauge-invariantly  (and  can be  
related to the ordinary pdfs renormalized 
in the minimal subtraction scheme ${\overline{\rm{MS}}}$~\cite{hef93})     
by going to  the high-energy pole in physical amplitudes~\cite{hef}.  
More general characterizations,  valid over the whole phase space,  
are  desirable,   and   currently the subject of much activity.  
Recent  results in     this area, see 
   e.g.~\cite{becher-neub,rogers,idilbisci,chered,bacch-diehl,rogers08,fhfeb07},  
    are likely  to     eventually   have a bearing    on forward jet physics.

In the next section we    discuss   the high-energy factorized form of the 
forward jet cross section. In Sec.~3 we  discuss applications to 
forward-central jet correlations.  We  give conclusions  in Sec.~4.

\section{Forward jet hadroproduction cross sections}
%\label{}
\nin
%%%%%%%%%%%%

The   presence of multiple   large-momentum scales  
in the  LHC forward jet  kinematics   
brings up  the  issue~\cite{ajaltouni,denterria,proceed09,michel_font}  of  whether    
fixed-order next-to-leading  calculations reliably describe the 
production process  or   significant contributions arise  beyond 
fixed  order  which call for 
perturbative QCD resummations.  
If   realistic  phenomenology   of      hadronic  jet final states 
  requires taking  into  account    at   higher  order     
 both logarithmic corrections   in  the large  rapidity  interval  
(of  BFKL  type)  
and logarithmic corrections  in  
the hard transverse momentum (of collinear type),    QCD  factorization at 
fixed transverse momentum can be used to achieve this~\cite{jhep09}.

\begin{figure}[htb]
\vspace{28mm}
\includegraphics{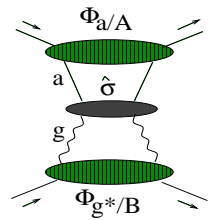}
\caption{ Factorized structure of the cross section. } 
\label{fig:sec2}
\end{figure}

The  k$_{\rm{T}}$-factorized form 
 of the  forward  jet     hadroproduction cross section  is  
     represented    in Fig.~\ref{fig:sec2}. 
Initial-state parton configurations  contributing to  
forward  jets   are asymmetric, 
with the parton in the top subgraph being  probed near  the mass shell and  
large   x,  
while  the parton in  the bottom subgraph is off-shell and small-x. 
The    jet  cross  section differential 
in the final-state   
transverse  momentum 
 $Q_t$  and  azimuthal angle $\varphi$ 
is given  schematically  by  
\begin{equation}
\label{forwsigma}
   {{d   \sigma  } \over 
{ d Q_t^2 d \varphi}} =  \sum_a  \int  \    \phi_{a/A}  \  \otimes \  
 {{d   {\widehat  \sigma}   } \over 
{ d Q_t^2 d \varphi  }}    \  \otimes \   
\phi_{g^*/B}    \;\; , 
\end{equation}
where  
$\otimes$ specifies  a convolution in both longitudinal and transverse momenta, 
$ {\widehat  \sigma} $  is the  hard scattering  cross section,  calculable 
 from  a  suitable off-shell continuation of 
perturbative   matrix elements~\cite{jhep09},  
$ \phi_{a/A} $ is the distribution of parton 
$a$ in hadron $A$ 
obtained from   near-collinear shower evolution, and $ \phi_{g^*/B} $ is  
  the gluon unintegrated distribution in hadron $B$ 
  obtained from non-collinear, 
  transverse momentum  dependent shower evolution.  

The multi-parton  matrix elements computed     in~\cite{jhep09}   
factorize,   in the high-energy limit,  
 not only in the collinear emission region but also  at finite angle.    
They  can be used    to take  
into     account    effects of coherence from multi-gluon emission  away 
from small angles,  
which become important for correlations among jets 
  across long      separations in rapidity.  We discuss this in the next section.

Note that  
  in the case of forward jet leptoproduction~\cite{mueproc90c} 
QCD factorization at fixed transverse momentum allows one to compute the 
high-energy asymptotic coefficients for  the coupling of forward jets to 
deeply inelastic scattering~\cite{hef,forwjetcoeff}. 
Since the early phenomenological studies~\cite{forwdis92}, forward jet 
leptoproduction has been investigated at Hera,  and 
will play a major role 
at  the proposed  future   lepton facilities~\cite{laycock} (LHeC, EIC). 
Measurements of 
 forward jet cross sections at   Hera~\cite{heraforw,aaron-hera} have illustrated that 
 neither fixed-order next-to-leading  
 calculations  nor  standard shower 
 Monte Carlo generators~\cite{heraforw,knut}, e.g.   \pythia\ 
  or \herwig,  are   able to   describe 
 forward jet  ep data.    
 (For   related  discussions of  central jet leptoproduction see~\cite{hj_ang}.) 
 Further   analyses of   ep data  and  physics at a future 
 high-energy lepton collider~\cite{laycock}  therefore  
    provide   additional  motivation for   developing    
 methods    capable of  
   describing   jet   production    beyond  the central rapidity region.

The approach above  can be combined with  parton showering  
in order to achieve a fully exclusive 
 description of the    final states associated  to  forward production.  To do this, 
 since Eq.~(\ref{forwsigma}) involves  off-shell 
 matrix elements encoding radiative effects beyond leading 
 order,  the basic point is that  one needs a    scheme 
 for merging consistently the hard radiation from  the 
short distance matrix element with the radiation from parton showering. 
 In~\cite{preprint}   the high-energy  factorization is used for  this  purpose.  
The other  important point  for the  coupling to parton showers is that 
because in the forward kinematics one of    the 
 longitudinal momentum fractions  x    in the  initial state 
    becomes small (see discussion around Fig.~\ref{fig:sec2}),      in order to take full  account of    
      multi-gluon emission    coherence  one needs to keep 
 finite-k$_{\rm{T}}$      terms     in the initial-state  parton 
      branching.  In the results shown in the next section this is done according to the 
      shower algorithm of~\cite{cascadedocu}. (See~\cite{gustafson,jadach09,watt_09} for 
      recent  work   on  related  methods.)

These features  of the merging and  showering 
distinguish this approach from  calculations in the BFKL picture, see e.g.~\cite{wallon10} at the 
next-to-leading order,  
in which  the parton   branching is taken to be  collinear.   In 
the picture of    Eq.~(\ref{forwsigma})   forward    jets  may   be 
produced  either   from the hard  
scatter subprocess or from the parton evolution subprocess.  
This differs from  purely  
collinear~\cite{michel_font}   or  BFKL~\cite{wallon10} approaches   in which   
   forward jets   are    produced by   hard matrix elements or impact factors.

%%%%%%%%%%%%
\section{Forward-central  jet correlations at the LHC}
%\label{}
\nin
%%%%%%%%%%%%

At the LHC it is possible to  
measure  events where jet transverse momenta 
p$_{\rm{T}} >  20$ GeV are produced several units of rapidity 
apart,    $\Delta \eta   \sim  3 \div 6$~\cite{grothe,ajaltouni}. 
Such multi-jet states can be relevant to 
new particle discovery processes as well as 
new aspects of standard model physics. 
Ref.~\cite{preprint}  investigates correlations  between  forward and  
 central jets,  in the   framework  discussed in the previous section, 
 examining the effects of   finite-angle gluon emission 
 across the large rapidity interval.  It  compares these with  
effects   of  the 
multi-parton interaction corrections  taken into account by~\cite{pz_perugia}.

The measurement of the 
azimuthal correlation of a central and forward jet     (Fig.~\ref{fig:jetcorr})  
provides a 
useful  probe of  how well   QCD multiple emissions  are described. 
  In~\cite{preprint} it is found that 
 while the average 
 of the  azimuthal separation  
  $\Delta \phi$ 
 between the  jets is not  affected very  much 
   as a function of 
 rapidity  by   finite-angle gluon emissions, 
 the detailed shape of the  $\Delta \phi$ distribution is. 

\begin{figure}[htbp]
\vspace{135mm}
\includegraphics{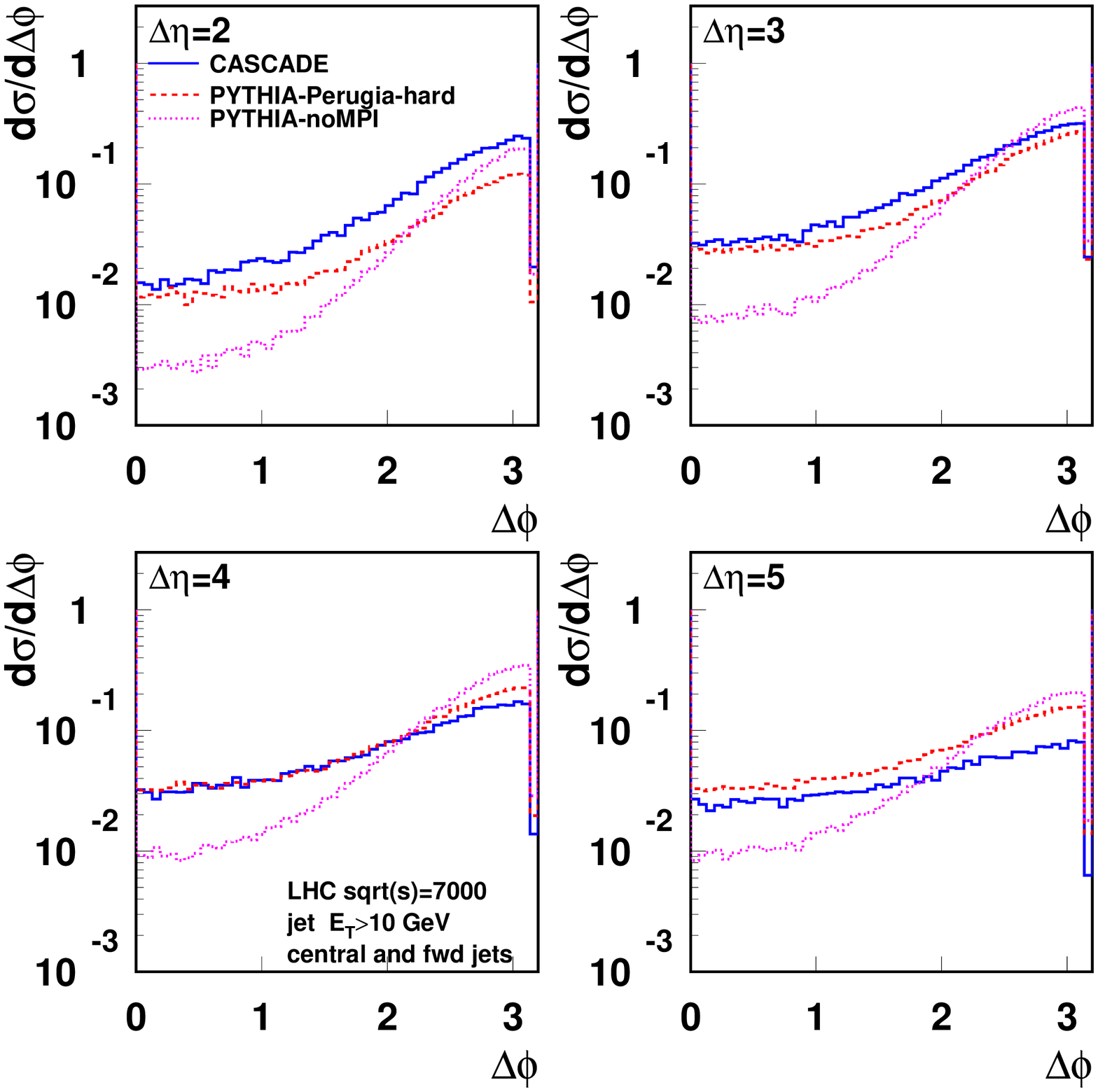}
\includegraphics{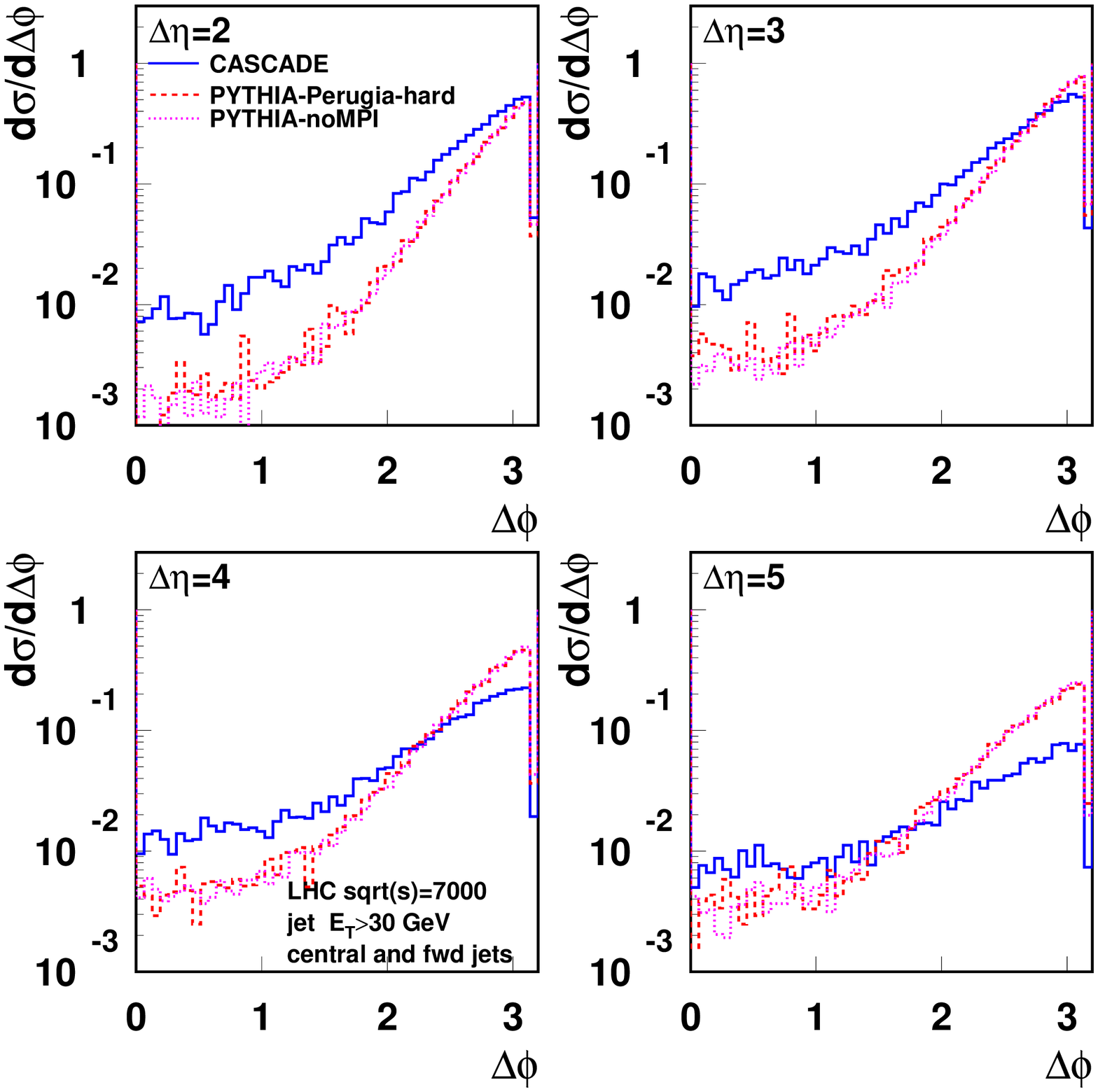}
\caption{ Cross section versus  azimuthal separation 
$\Delta \phi$  between  central and  forward jet,  at  different  
rapidity  separations $\Delta \eta$,  for jets with transverse energy $E_T > 10$ GeV (top)  and 
$E_T > 30$ GeV (bottom)~\cite{preprint}. } 
\label{fig:azsigma}
\end{figure}

The cross section as a function of the 
azimuthal separation $\Delta \phi$  between  central and  
forward jets reconstructed  with the      Siscone algorithm~\cite{fastjetpack} ($R =  0.4$)
  is shown in Fig.\ref{fig:azsigma}~\cite{preprint} for different  rapidity separations. 
The solid blue  curve is the prediction based on   implementing  the factorization~\cite{jhep09}  
of Eq.~(\ref{forwsigma})   in the  parton-shower event generator~\cite{cascadedocu} 
(\protect\CASCADE); the red and purple curves are  
 the predictions  based on calculations with 
  collinear  parton-showering~\cite{pz_perugia} (\protect\PYTHIA), respectively   
  including  multiple interactions  and without multiple interactions.  
 
The decorrelation as a function of $\Delta\eta$ increases in \CASCADE\ as well as in \PYTHIA . 
In the low $E_T$  region (Fig.~\ref{fig:azsigma} (top)) the 
increase in decorrelation with increasing $\Delta\eta$ is    significant. 
The cross section for jet separation up to $\Delta\eta < 4 $ is 
 similar between \CASCADE\ and \PYTHIA\  with multiparton interactions, whereas a clear difference is seen to \PYTHIA\ without multiparton interactions. However, at large $\Delta\eta > 4 $ the decorrelation predicted by \CASCADE\ is significantly larger than the prediction from  multiparton interactions. 
In the higher $E_T$ region    
(Fig.~\ref{fig:azsigma} (bottom))
\CASCADE\ predicts everywhere a larger decorrelation. In this region  the influence of multiparton interactions in \PYTHIA\ is small and the difference to \CASCADE\ comes entirely 
from the different parton shower. 

\begin{figure}[htbp]
\vspace{79mm}
\includegraphics{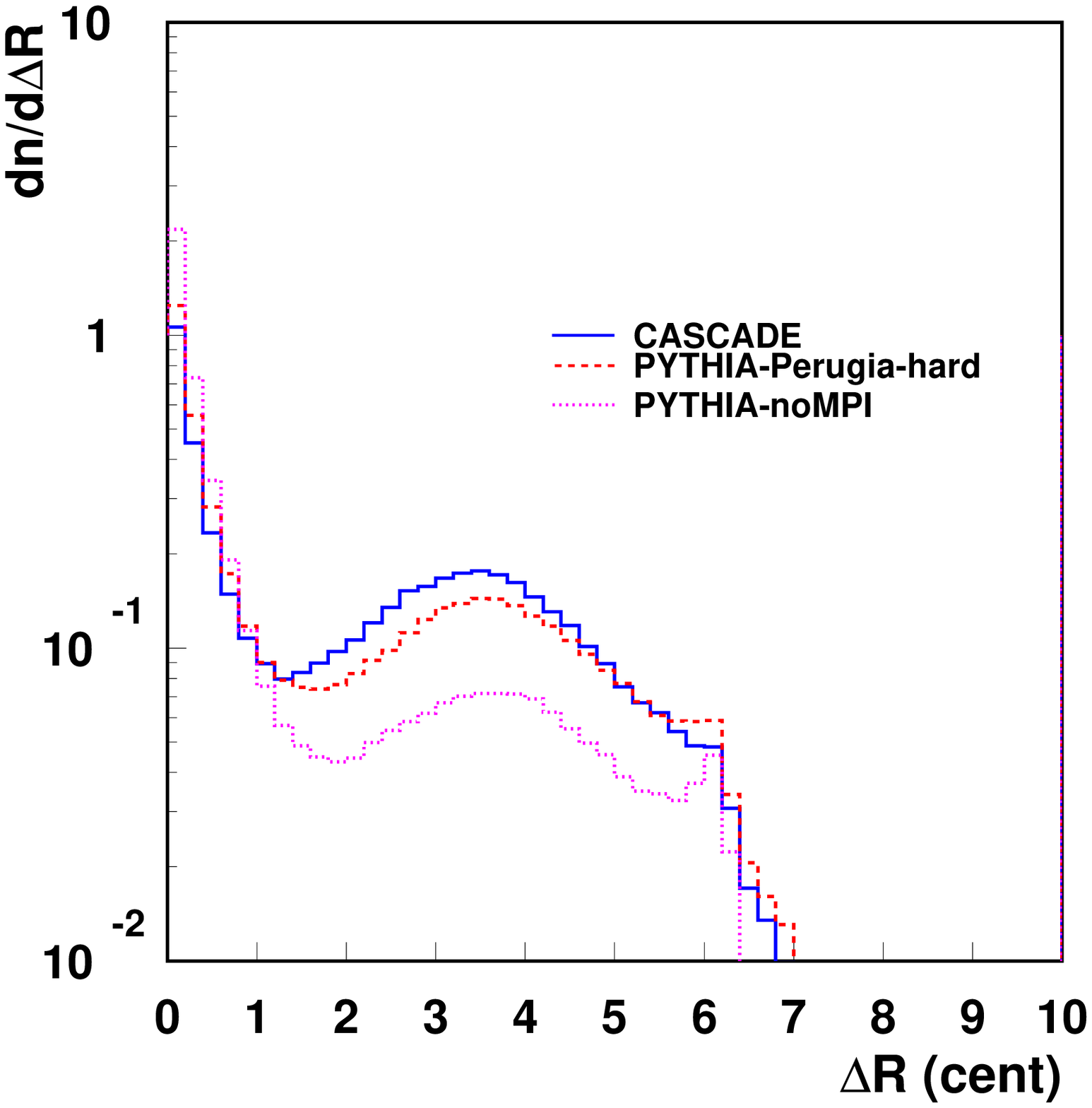}
\includegraphics{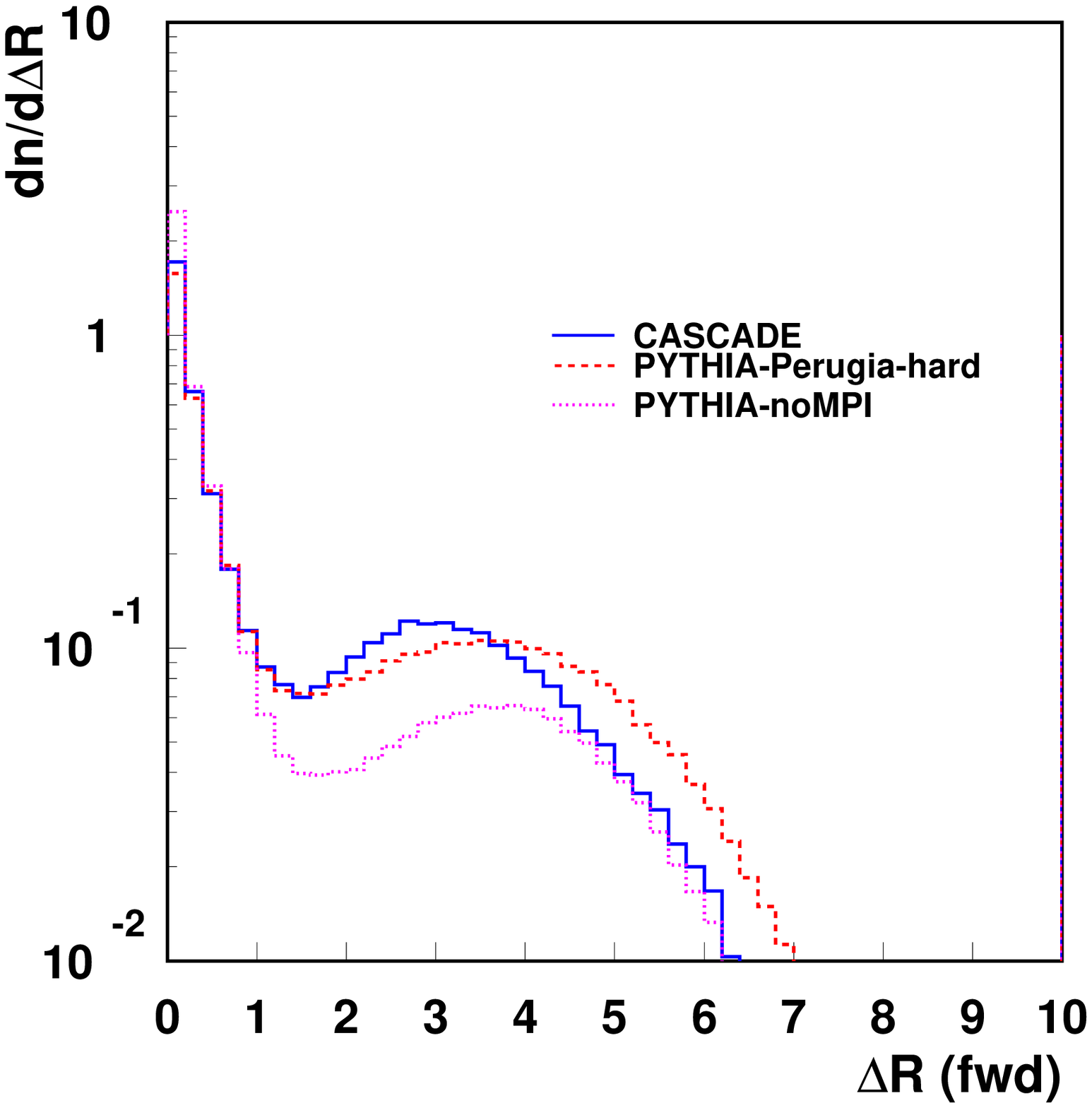}
\includegraphics{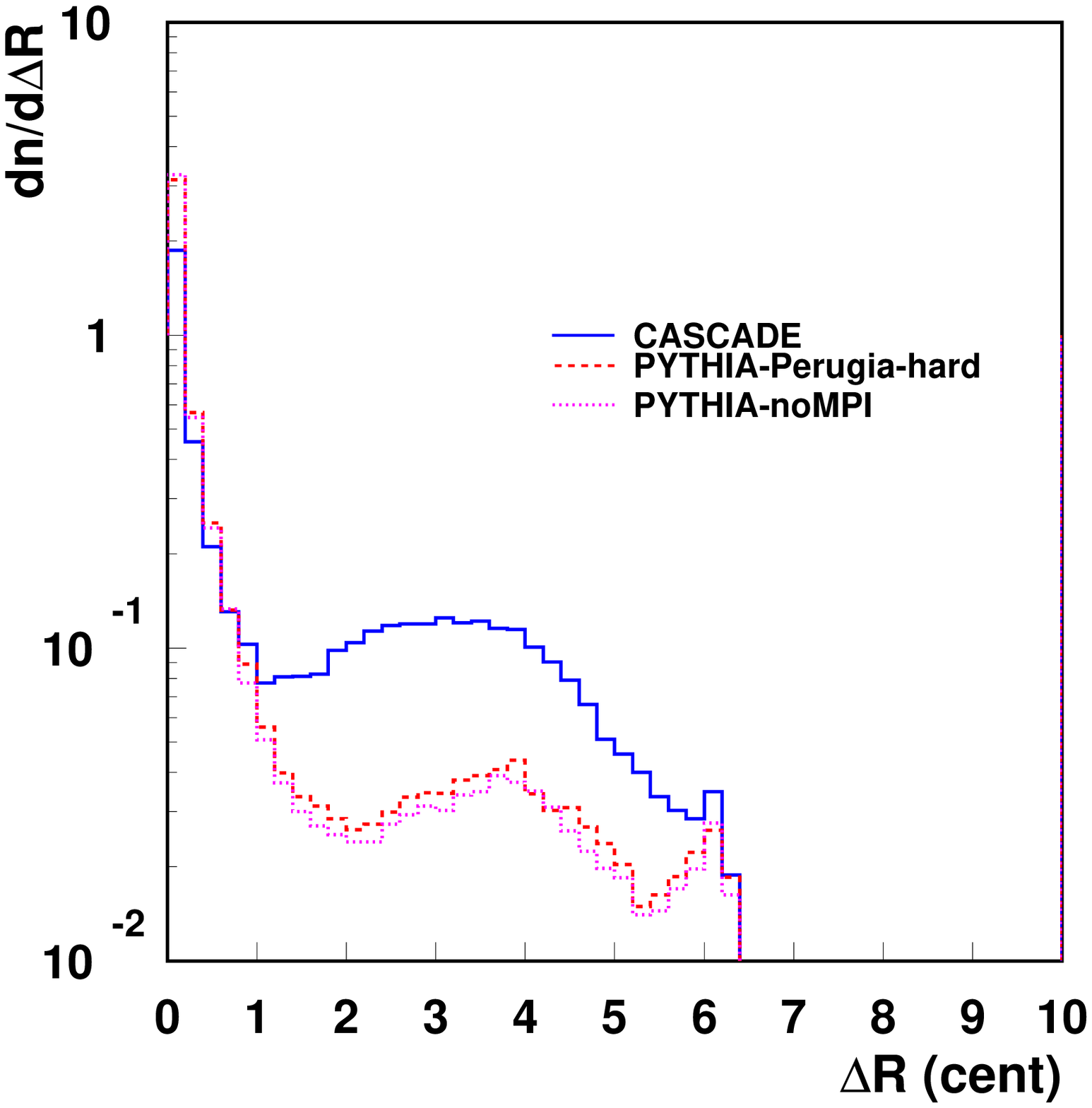}
\includegraphics{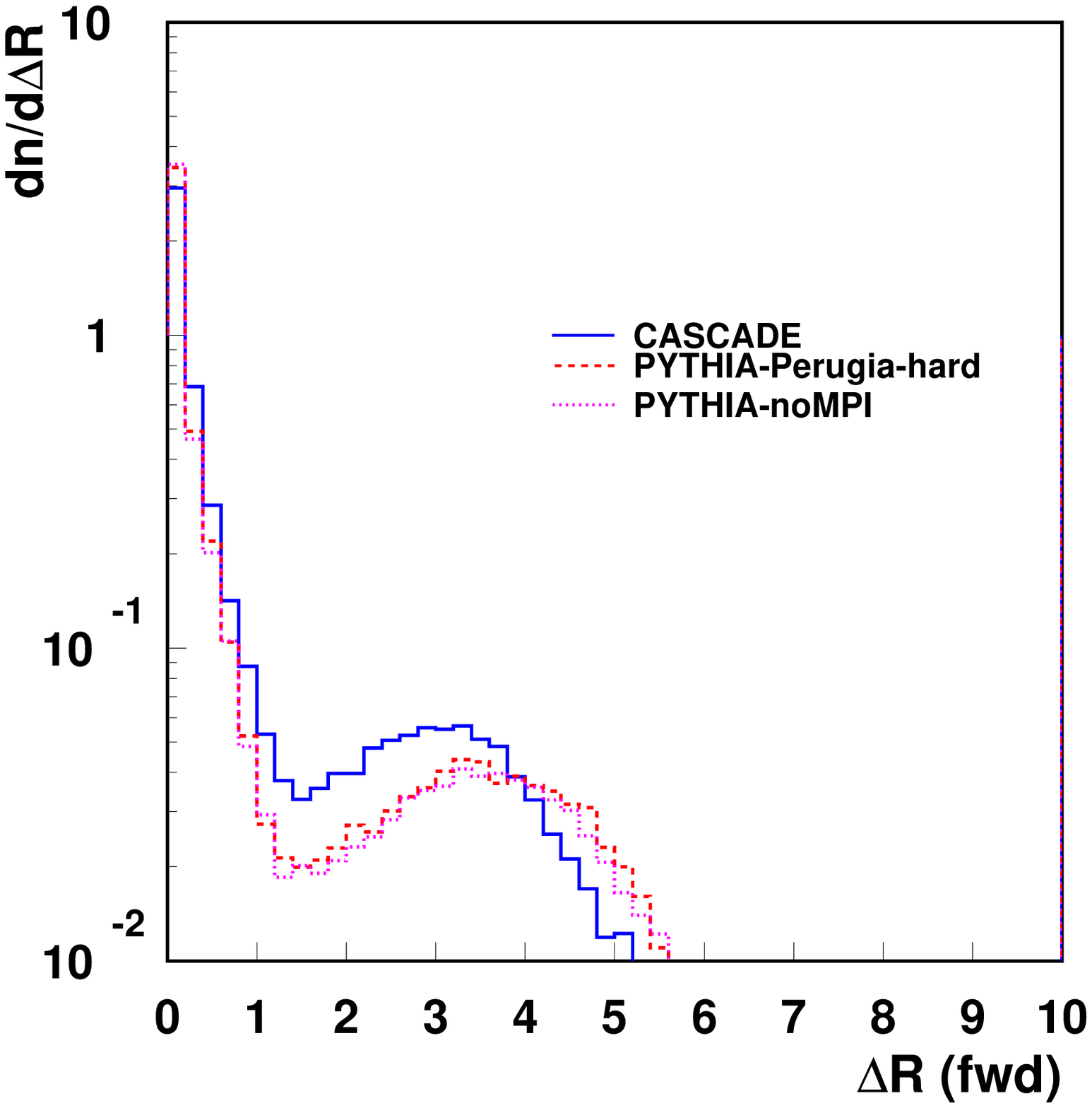}
\caption{ $\Delta R$ 
distribution of the central  ($|\eta_c|<2$, left)  and forward jets ( $3 < |\eta_f| <5 $, right) 
 for $E_T > 10$~GeV (upper row) and  $E_T>\!30$~GeV (lower row)~\cite{preprint}.} 
\label{fig:deltar}
\end{figure}

Distinctive  effects from  the high-energy, noncollinear corrections to   parton showers  
are also observed (Fig.~\ref{fig:deltar}~\cite{preprint})  in  the 
$\Delta R =   \sqrt{\Delta \phi^2 + \Delta \eta^2 }$ distribution, where 
 $\Delta \phi= \phi_{jet} - \phi_{part}$ ($\Delta \eta= \eta_{jet} - \eta_{part}$) is the azimuthal 
 (rapidity) difference between the jet and the 
 corresponding parton from the matrix element.      This distribution probes to what extent 
 jets are dominated by hard partons in the matrix element  or  receive 
 significant contributions  from the  showering. The  large-$\Delta R$  region  is  seen     
 to be enhanced by noncollinear corrections, and while this signal can be mimicked by 
 multi-parton interactions  for   low $E_T$ jets, this  no longer applies  as $E_T$  
  increases.

%%%%%%%%%%%%%%%%
\section{Conclusion}
\nin
%%%%%%%%%%%%%%%%
Jet  physics   in the forward region  at hadron-hadron colliders     is   a largely new   area of   
experimental and theoretical  activity, and  enters   the 
LHC  program   in  both     new particle discovery 
processes (see e.g.  vector boson fusion channels~\cite{vbf-cms,vbf-atlas} for Higgs boson 
searches) and  new aspects of 
standard model physics (e.g.,   QCD at small x and its interplay with cosmic ray physics, 
see~\cite{denterria,engel}).     
In this kinematic 
 region the evaluation of QCD theoretical predictions  is made complex 
due to  the presence of multiple mass scales, and the question arises 
  of whether   
perturbative QCD resummations  and/or   corrections   from multiple  parton interactions   
are called for 
in order  to go beyond the case of central jets~\cite{centr-jet-atl,centr-jet-cms}. 
  
The  factorization~\cite{jhep09}  allows one to sum consistently  to all perturbative 
orders both large logarithms of rapidity and 
 large logarithms of transverse momentum.   
 Based on this analysis,     
  contributions     to the QCD parton cascades   
from finite-angle multi-gluon emission~\cite{preprint}    
 over wide rapidity intervals    are found to 
affect significantly the predictions for forward jets.  

Distinctive effects are seen in  particular by  considering  correlations between forward and 
central  jets~\cite{preprint},    e.g.   azimuthal correlations.
 Phenomenological studies based on  measurements of  these correlations will be relevant 
 for  tests of   initial state radiation and 
for the QCD tuning of  Monte Carlo event  generators. 
   (For the counterpart of this in the case of central jet pairs see  the first 
LHC measurements~\cite{centr-jet-atl,centr-jet-cms}.)     They can also  be relevant 
to gain  better control on   the  structure of the  final states   associated  with 
  heavy  particle production  
(e.g.,  underlying jet  activity  in scalar boson production~\cite{deak_etal_higgs}). 

This analysis can be extended to the case  of  forward and backward jets. It  
 can thus serve 
to estimate the size of backgrounds  from  QCD radiation 
  in Higgs searches      from vector boson fusion~\cite{vbf-cms,vbf-atlas}.

Studies of  forward     high-p$_{\rm{T}}$  production, 
  such as those discussed in this  
  article,   will be complemented at  later  stages  of the LHC program 
  by studies in  other areas of 
  forward physics employing   near-beam proton taggers~\cite{albrow_review}. Both 
  the   high-p$_{\rm{T}}$ and proton-tagging  measurements can  contribute to 
either   standard-candle  or discovery physics.   In addition, both will provide 
inputs  on (hard and soft)  forward particle production that 
will serve   for   the modeling of high-energy air showers~\cite{engel} in cosmic 
ray experiments.

%%%%%%%%%%%%%%%%%%%%%%%%%%%
\section*{Acknowledgments}
\nin
%%%%%%%%%%%%%%%%  
  The results presented in this 
 article   have been obtained   in collaboration with 
M.~Deak, H.~Jung and K.~Kutak. 


%% References


%\bibliographystyle{elsarticle-num}





% \end{thebibliography}

%%%%%%%%%%%%%%%%%%%%
%\vfill\eject

\input{bib_wtup_mont10}

\end{document}

%% file: bib_wtup_mont10.tex
%%%%%%%%%%%%%%%